\documentclass[showpacs, nofootinbib, aps, prd, superscriptaddress,10pt, showkeys, notitlepage,twocolumn]{revtex4-1}

\usepackage{graphicx,amssymb,amsmath,amsthm,amsfonts,epsfig}

\usepackage[linktocpage]{hyperref}
\usepackage[usenames,dvipsnames]{color}
\usepackage{epstopdf}
\definecolor{darkred}{rgb}{0.5,0,0}
\definecolor{darkgreen}{rgb}{0,0.5,0}
\definecolor{darkblue}{rgb}{0,0,0.5}
\definecolor{prussian}{rgb}{0.0, 0.19, 0.33}
\definecolor{richelectricblue}{rgb}{0.03, 0.57, 0.82}
\definecolor{teal}{rgb}{0.0, 0.5, 0.5}
\definecolor{mediumseagreen}{rgb}{0.24, 0.7, 0.44}
\definecolor{lust}{rgb}{0.9, 0.13, 0.13}
\definecolor{ballblue}{rgb}{0.13, 0.67, 0.8}
\definecolor{darkcyan}{rgb}{0.0, 0.55, 0.55}
\definecolor{mountainmeadow}{rgb}{0.19, 0.73, 0.56}
\definecolor{palecarmine}{rgb}{0.69, 0.25, 0.21}
\definecolor{richcarmine}{rgb}{0.84, 0.0, 0.25}
\definecolor{tangelo}{rgb}{0.98, 0.3, 0.0}
\definecolor{venetian}{rgb}{0.784,0.031,0.082}
\definecolor{bdfrance}{rgb}{0.192,0.549,0.906}

\newcommand{\bea}{\setlength\arraycolsep{2pt} \begin{eqnarray}}
\newcommand{\eea}{\end{eqnarray}}

\hypersetup{colorlinks=true, citecolor=venetian,
linkcolor=bdfrance, urlcolor=lust}
\usepackage{amsmath,amssymb}
\usepackage{tensor}
\usepackage{mathtools}
\usepackage{amsbsy}
\usepackage{bm}

\newcommand{\be}{\begin{equation}}
\newcommand{\ee}{\end{equation}}
\newcommand{\bear}{\begin{eqnarray}}
\newcommand{\eear}{\end{eqnarray}}

\newcommand{\rp}{r_{+}}
\newcommand{\rmi}{r_{-}}
\newcommand{\tM}{\tau_{\rm M}}
\newcommand{\tP}{\tau_{\Phi}}

\def\apj{{ApJ}}

\def\apjl{{ApJL}}

\def\aap{{A\&A}}

\def\mnras{{MNRAS}}

\def\prd{{Physical Review D}}

\def\04a{{2004 a}}
\def\04b{{2004 b}}

\begin{document}

\title{Multipole moments do not uniquely characterize spacetimes beyond general relativity}

\date{\today}

\author{Arthur G. Suvorov}\thanks{arthur.suvorov@tat.uni-tuebingen.de}
\affiliation{Theoretical Astrophysics, IAAT, University of T{\"u}bingen, T{\"u}bingen, D-72076, Germany}
\author{George Pappas}\thanks{gpappas@auth.gr}
\affiliation{Department of Physics, Aristotle University of Thessaloniki, Thessaloniki 54124, Greece}

\begin{abstract}
\noindent Spacetimes in general relativity can be uniquely decomposed into a set of multipole moments. Given the usefulness of moments in the categorization of radiation patterns, tidal deformations, and other phenomena associated with compact objects, a number of studies have explored their construction in beyond-Einstein theories of gravity. It is shown here that uniqueness does not necessarily extend across theories: by comparing a few static and spherically-symmetric solutions in different theories, we find that two distinct objects can possess the same Geroch-Hansen moments. Moreover, two metrics can match and yet take different moments. Implications of this result are explored in the context of black-hole shadows and ``universal'' relations hinging on moment computations.
\end{abstract}
  
\maketitle


\section{Introduction} \label{sec:intro}

Multipole moments comprise a tool for categorising solutions to differential equations. By making use of projections onto a spectral basis, information pertaining to a given solution can be compressed into a set of coefficients for ordered eigenfunctions. The classical example is that of the Poisson equation with Dirichlet boundary conditions, solutions to which can be uniquely decomposed into a set of spherical harmonics multiplied by increasingly negative powers of radius \cite{ll71}. This equation arises in the study of Newtonian gravity and Maxwellian electrostatics, and multipole moments are routinely used in the description of the gravitational \cite{rox01,anita08,laar99,Pappas:2012ns} and magnetic \cite{greg11,gast12,sm20} fields within and surrounding celestial bodies.

Defining moments in general relativity (GR) is more complicated. Several methods have been introduced, though most essentially involve rewriting the field equations in such a way that Poisson or wave equations arise for appropriate variables. For example, \citet{thorne80} employed the de Donder gauge to recast the Einstein equations into wave equations for a rescaled metric such that a solution can be decomposed into harmonic projections of the relevant Green's function, thereby allowing the moments again to be read off from Laurent coefficients in a special coordinate system. In a more geometric scheme, Geroch \cite{ger70a,ger70} and Hansen \cite{han74} showed that Poisson equations for some particular potentials arise from the stationary, vacuum Einstein equations projected onto spacelike hypersurfaces. Owing to diffeomorphism invariance, moment decompositions are particularly powerful in GR as they allow for an unambiguous interpretation of spacetime parameters. Imagine encountering the Kerr solution in unfamiliar coordinates with relabelled constants: identifying the mass and spin may be an arduous task without an invariant expansion at hand (see, e.g., Refs.~\cite{quev90,hern94,heus96} for discussions). Moments are also convenient for characterising gravitational radiation \cite{isac68,isac68b,thorne80,comp18} and the visibility of astrophysical sources \cite{ryan95,ryan97,aasi14,smg16,skk24,asp25}.

Although the procedure for computing moments is not unique in GR, it has been established that a given set of moments {determines a vacuum metric up to isometries} and the prescriptions considered in the literature are equivalent within the appropriate realms of validity \cite{xanth79,bs81,sb83,bs00}. 
Notably, \citet{gurs83} showed that the Geroch-Hansen moments are equivalent to the Thorne moments for stationary systems (see also Ref.~\cite{voor25} for some refinements with respect to proof gaps).
However, the goal of this paper is to demonstrate that dropping the qualifier ``in GR'' renders the more-general statement \emph{untrue}. 

We consider a few example, asymptotically-flat solutions in some scalar-tensor theories of gravity and compute their moments using a Geroch-Hansen procedure. While the line elements and scalar profiles differ, it is shown that their moments can match; this implies that knowledge of the latter is insufficient to reconstruct a spacetime in the absence of a fixed theory. This result appears to have gone unpenned in the literature, though could have important consequences. For example, ``universal relations'' involving quadrupole moments and other intrinsic parameters of compact objects have been extensively explored in a variety of contexts \cite{yagi13,Pappas:2013naa,Yagi:2014bxa,yagi14,mart14,pap19,fil24}. If, however, the moments themselves vary \emph{across} theories, it becomes less clear how one may utilise these relations unless one \emph{a priori} assumes a fixed theory of gravity when comparing predictions to observational data. {We emphasise at this point that we do not address the harder problem of when a given set of moments uniquely corresponds to solutions \emph{within a fixed theory}. Such a problem boils down to assessing whether Birkhoff-like theorems hold, which generally requires case-by-case handling \cite{ghosh25}; some first steps towards a result are presented in Appendix~\ref{sec:Appendix}.}

This work is organised as follows. Section~\ref{sec:spacetimes} introduces the general form of the spacetimes we consider, for which the moments are defined in Sec.~\ref{sec:multmoms}. Some specific examples are considered in Section~\ref{sec:examples}, where the (non-zero) moments are computed.  Section~\ref{sec:degen} makes use of these expressions to validate the claim made in the paper title; an application to black hole shadows is given in Sec.~\ref{sec:deflection}. Closing discussion and outlook is provided in Section~\ref{sec:discussion}.

\section{Spacetime background} \label{sec:spacetimes}

We focus on static and spherically-symmetric spacetimes. In Schwarzschild-like coordinates $\{t,r,\theta,\varphi\}$, we consider a line element of the form
\begin{equation} \label{eq:genline}
ds^2 = -A(r) dt^2 + B(r) dr^2 + D(r) \left(d \theta^2 + \sin^2\theta d \varphi^2 \right),
\end{equation}
for some functions $A$, $B$, and $D$ behaving appropriately to ensure asymptotic flatness. In addition to the metric \eqref{eq:genline}, we allow for the existence of scalar, $\phi$, and electric\footnote{Electromagnetic duality implies we could equally talk about magnetic moments without loss of generality \cite{hoen76}.}, $E$, potentials which respect the symmetries of the spacetime. Specific examples, corresponding to select scalar-tensor theories of gravity, are considered in Sec.~\ref{sec:examples}. 

\subsection{Geroch-Hansen multipole moments} \label{sec:multmoms}

Here we recap the construction of the Geroch-Hansen moments. Because spherical harmonics multiplied by inverse powers of radius define the eigenbasis for the Laplace operator, \emph{Newtonian} moments can be simply read off from the coefficients of the gravitational potential once projected into harmonics \cite{ger70a}. By way of analogy, \citet{ger70} and \citet{han74} recast the vacuum Einstein equations, $R_{\mu \nu} = 0$, into conformally-invariant Laplace or Poisson ones, respectively, in order to identify a set of moments describing the geometry outside of a stationary object. In particular, the assumption of \emph{asymptotic flatness} implies that spacelike slices can be completed with a point at infinity, $\Lambda$; \emph{stationarity} guarantees that slices are isomorphic so that only one need be considered, $S$, and \emph{conformal invariance} means that physical information extracted at $\Lambda$ is unaffected by redefinitions of clocks and rulers (i.e. Weyl transformations; see Refs.~\cite{chru89,heus96} for more technical discussions).

In principle, the Geroch-Hansen procedure is not special to GR or the metric specifically provided appropriate potentials can be identified  \cite{sm16,cano22}. For instance, the equations of motion for $\phi$ in a minimally-coupled scalar-tensor theory can also be rewritten as a Laplace-Beltrami equation \cite{ps15}.  An important feature of the moments for \emph{spherically symmetric} spacetimes \eqref{eq:genline} is that only the monopole moments may be nonzero since coefficients of angular harmonics trivially vanish.

Associated with a stationary spacetime is the existence of a timelike Killing vector, $\boldsymbol{\xi}$. From this one can introduce its norm\footnote{One may further introduce a twist associated with $\boldsymbol{\xi}$ to define a set of current multipole moments characterising the spacetime angular-momentum distribution, though this is unnecessary for our purposes. Note that Greek indices run over spacetime variables while Latin ones are reserved for purely spatial components.}, $\lambda = \xi^{\alpha} \xi_{\alpha}$, which is essentially the lapse function. It can be used to foliate the spacetime into a set of spacelike 3-slices over which the relevant equations are subsequently introduced and solved to deduce the moments. Indeed, the necessity of stationarity is tied to the well-known problem of defining mass in dynamical spacetimes so that only one slice need be considered, with metric $\boldsymbol{h}$ defined via [compare equation \eqref{eq:genline}]
\begin{equation}
ds^2 = \lambda dt^2 - \lambda^{-1} h_{ij} dx^{i} dx^{j}.
\end{equation}

Consider a triplet of fields $\tM$, $\tP$, and $\tau_{\rm P}$ as functions of $\lambda$, $\phi$, and $E$ respectively. These are chosen to satisfy ($A = M,\Phi,P$ for mass, scalar, and electric moments)
\begin{equation} \label{eq:multmomeqn}
\left(D^{i} D_{i} - \frac{\mathcal{R}}{8} \right) \tau_{A} = 0,
\end{equation}
where $D^{i}$ and $\mathcal{R}$ are the covariant derivative and Ricci scalar on $(S,\boldsymbol{h})$, respectively. In GR and potential-free scalar-tensor theories, equation \eqref{eq:multmomeqn} can be solved identically\footnote{We work with Laplace-like equations rather than the Hansen-like generalisations with a right-hand side \cite{han74}. As discussed by \citet{sm16} and references therein, the same moments are recovered in either case and, as solving the equations of motion is more difficult in non-GR theories, the computational advantage of Hansen's adjustment is often moot.} by choosing the $\tau_{A}$ appropriately \cite{ps15}. In general, however, this is not possible for an arbitrary set of field equations and a case-by-case approach must be used to solve equation \eqref{eq:multmomeqn}, numerically or otherwise \cite{sm16,cano22}. 

Given potentials $\boldsymbol{h}$, $\phi$, and $E$, one can solve equation \eqref{eq:multmomeqn} to determine the moments using the following recipe \cite{ger70a,ger70,han74}:
\begin{itemize}
\item[(i)]{Determine the $\tau_{A}$ as functions of the appropriate variables ($\lambda$, $\phi$, and $E$, or the twist and its generalisations in stationary cases \cite{sm16}). These equations have to be solved subject to boundary conditions which reproduce the GR solutions when $R_{\mu \nu} = 0$.}
\item[(ii)]{Choose a new radial coordinate, $\bar{R}(r)$, such that the point at infinity ($\Lambda$) is mapped to a finite radius. This step is not strictly necessary, but simplifies the identification of a suitable conformal factor \cite{han74}.}
\item[(iii)]{Choose a conformal factor $\Omega$ (required to satisfy $\Omega =0, \tilde{D}_{a} \Omega = 0$, and $\tilde{D}_{a} \tilde{D}_{b} \Omega = c h_{ab}$ for some $c$) to carry out the conformal completion with $\tilde{\boldsymbol{h}} = \Omega^2 \boldsymbol{h}$. For the simplest treatment, one picks this function such that $\tilde{\boldsymbol{h}} = \boldsymbol{\delta}$ for Euclidean 3-metric. Since equations \eqref{eq:multmomeqn} are conformally invariant, flat-space analogies may now be directly applied on the conformally-completed space, $\tilde{S} = S \cup \Lambda$ \cite{ger70a}.}
\item[(iv)]{The monopole moments read $A_{0} = \lim_{\bar{R} \to \Lambda} \Omega^{-1/2} \tilde{\tau}_{A}$ for conformally-transformed potentials $\tilde{\tau}_{A}$ where $A = M,\Phi,P$. (Higher-order moments are similarly computed by applying derivative operators to $\tilde{\tau}_{A}$ to isolate the correct coefficient of the relevant tensor harmonic).}
\end{itemize}

Following Ref.~\cite{sm16}, for the mass moments we impose the following boundary conditions  on equation \eqref{eq:multmomeqn}
\begin{equation} \label{eq:massbc}
\tM(1) = 0, \tM'(1) = -1/2,
\end{equation}
where the prime denotes differentiation with respect to argument. The choices made in \eqref{eq:massbc} ensure that the Geroch solution (up to sign\footnote{The choice of sign in these conditions is conventional. Imposing instead $\tM'(1) = 1/2$ yields moments of the opposite sign \cite{ger70}.}), $\tM = \sqrt{\lambda} -1$, is recovered in the source-free, GR limit ($R_{\mu \nu} = 0$) where equation \eqref{eq:multmomeqn} reduces to $0=\tfrac{1}{2} \lambda^{-1} \tau_{\rm M}'(\lambda) + \tau_{\rm M}''(\lambda)$ (see equation 19 in Ref.~\cite{sm16}). For the scalar moments, we set
\begin{equation} \label{eq:scalarbc}
\tP(1) = 0, \tP'(1) = 1,
\end{equation}
to recover the definitions of Ref.~\cite{ps15} in the purely scalar-tensor limit (see their expression 34). Provided the electromagnetic sector is minimally coupled, the definition of the vector moments are unaltered with respect to that of GR and the results of Refs.~\cite{simon84,fod89,hoen90} can be applied directly to a Coulomb potential. We refer the reader to Refs.~\cite{sot04,fod21} for a discussion on computational elements.

\section{Moments across different theories of gravity} \label{sec:examples}

We now turn to the main results of this work: computing the moments for a few spacetimes in different theories of gravity to facilitate a direct comparison. 

\subsection{Reissner-Nordstr\"{o}m} \label{sec:rn}

We first consider the classical Reissner-Nordstr\"{o}m (RN) solution from GR. For this spacetime, we have 
\begin{equation} \label{eq:rnmet}
A(r) = 1 - \frac{2 M}{r} + \frac{Q^2}{r^2},
\end{equation}
$B(r) = 1/A(r)$, and $D(r) = r^2$, where one identifies $M$ and $Q$ within expression \eqref{eq:rnmet} as the mass and electric charge, respectively. As expected, these constants are precisely the mass and electric monopoles \cite{fod89}, viz.
\begin{equation} \label{eq:rnmoments}
M_{0} = M, P_{0} = Q,
\end{equation}
where, obviously, the scalar monopole vanishes ($\Phi_0 = 0$). All other moments vanish thanks to spherical symmetry.

\subsection{Charged Janis-Newman-Winicour}  \label{sec:jnw}

Consider a massless scalar field minimally-coupled to GR through the action
\begin{equation} \label{eq:stactionein}
S=\frac{1}{16\pi}\int d^{4} x \sqrt{-g} \left(R-2 \nabla^\mu \phi \nabla_\mu \phi - F_{\alpha \beta} F^{\alpha \beta} \right),
\end{equation}
where $\boldsymbol{\nabla}$ is the 4-connection and $F_{\mu \nu} = \partial_{\mu} A_{\nu} - \partial_{\nu} A_{\mu}$ is the Faraday tensor. In the limit of vanishing electromagnetic charge ($E = 0$), the unique solution describing the geometry outside of a static and spherically-symmetric object is known as the Janis-Newman-Winicour (JNW) solution (though, historically, it was first discovered by \citet{fisher99}). The solution was recently generalised by \citet{tur21} to include a radial electric field, $E^{r} = \sqrt{-F_{rt} F^{rt}}$ with $A_{\mu} = (A_{t},0,0,0)$. The metric functions for the charged JNW spacetime read
\begin{equation} \label{eq:stmet}
\begin{aligned}
A(r) &= \left[ \frac{ \rp \left( \frac{r - \rmi}{r - \rp}\right)^{n/2} - \rmi \left( \frac{r - \rp}{r - \rmi} \right)^{n/2}} {\rp - \rmi} \right]^{-2} \\
&= 1 - \frac{2M}{r} +\frac{2 M^2 (n-1) + n Q^2}{r^2} + \mathcal{O}\left({\frac{1}{r^3}}\right),
\end{aligned}
\end{equation}
$B(r) = 1/A(r)$, and
\begin{equation} \label{eq:dfunc}
D(r) =  \frac{\left(r - \rp\right) \left( r - \rmi \right)}{A(r)},
\end{equation}
where $n r_{\pm} = M \pm \sqrt{M^2 - Q^2}$ so that $n$ effectively sets the amplitude of the scalar field through
\begin{equation} \label{eq:phist}
\begin{aligned}
\phi(r) &= \frac{\sqrt{1 - n^2}}{2} \log \left( \frac{r - \rp}{r - \rmi} \right)\\
&= -\sqrt{M^2 - Q^2} \frac{\sqrt{1/n^2 - 1}}{r}  +\mathcal{O}\left( \frac{1}{r^2} \right).
\end{aligned}
\end{equation}
If we relabel $n=M/l$, we recover expression (63) from Ref.~\cite{ps15} in the $Q \to 0$ limit (i.e., the JNW solution \cite{buch59}). While $r\to r_{\pm}$ correspond to curvature singularities and the scalar field diverges there as well, the construction serves as a proof-of-concept: the metric could describe the geometry surrounding a hairy, exotic object \cite{hertz25}.

As found by \citet{ps15} in the $Q \to 0$ limit, the non-zero multipoles are the mass monopole moment, $M_{0} = M$, and the scalar monopole moment, $\Phi_{0}(Q=0) = -M \sqrt{1/n^2 - 1}$. More generally though, we see that the leading-order ($1/r$) behaviour in the scalar field \eqref{eq:phist} is affected through a simple rescaling of the mass, and thus $\Phi_{0} = -\sqrt{M^2 - Q^2} {\sqrt{1/n^2 - 1}}$. Given the minimally-coupled nature of the electromagnetic sector in the action \eqref{eq:stactionein}, the electric moments follow from the GR case where the Ernst potential acquires an additive correction proportional to the Coulomb potential \cite{hoen90}. This can be seen from the electric field itself:
\begin{equation}
    E^{r}(r) = \frac{Q}{r^2} + \frac{Q \left(1-n\right) \left(r_{-} + r_{+}\right)}{r^3} + \mathcal{O}\left(\frac{1}{r^4}\right).
\end{equation}
As such, the only non-zero electric moment is $P_{0} = Q$.

\subsection{Bocharova-Bronnikov-Melnikov-Bekenstein} \label{sec:bbmb}

\begin{table*}
	\caption{Key properties and moments for the solution families considered in this work. For each metric, we have $B(r) = 1/A(r)$.
}
\begin{tabular}{l|cc|ccc}
	    \hline
	    Metric family & $A(r)$ & $D(r)$ & Scalar monopole & Electric monopole & Mass monopole \\
	    	    \hline
	    	    	    \hline
Reissner-Nordstr\"{o}m & Eq.~\eqref{eq:rnmet} & $r^2$  & 0 & $Q$ & $M$ \\ 
Charged Janis-Newman-Winicour & Eq.~\eqref{eq:stmet} &  Eq.~\eqref{eq:dfunc} & $-\sqrt{M^2 - Q^2} {\sqrt{1/n^2 - 1}}$ & $Q$ & $M$ \\ 
Bocharova-Bronnikov-Melnikov-Bekenstein & Eq.~\eqref{eq:bbmb} & $r^2$  & $\phi_{\infty}\sqrt{M^2-Q^2}  $ & $Q$ & $M$ \\ 
     \hline
	    \hline
	\end{tabular}
	\label{tab:keyprops}
\end{table*}

Consider now a theory where the Einstein-Hilbert action is \emph{conformally coupled} to a scalar sector through 
\begin{equation} \label{eq:BBMBaction}
S=\int d^4 x\sqrt{-g} \left( \frac{R}{2 \kappa} - \frac{1}{2} \nabla^\mu \phi \nabla_\mu \phi  -\frac{1}{12} R\phi^2 \right),
\end{equation}
where $\kappa$ is set by the scalar asymptotics, $\phi_{\infty}=\pm {\sqrt {6/\kappa}}$.  

The so-called Bocharova-Bronnikov-Melnikov-Bekenstein (BBMB) \cite{bbm70,bek74} black hole is an exact solution in the theory described by \eqref{eq:BBMBaction}. Although the general solution is represented by a three-parameter family, one of them is forcibly eliminated by the requirement that the solution admits a smooth horizon and no naked singularities. With this restriction, it becomes the \emph{unique} static and spherically-symmetric vacuum solution in the conformally-coupled theory \eqref{eq:BBMBaction} \cite{xanth91}. While known to be unstable to polar-parity perturbations \cite{bron78,kob14}, calling into question the health of the theory, it again provides us with an avenue for proof-of-concept. The BBMB metric functions read
\begin{equation} \label{eq:bbmb}
A(r) =  1-\frac{2M}{r} + \frac{M^2}{r^2},
\end{equation}
with $B(r) = 1/A(r)$ and $D(r) = r^2$. The scalar field is 
\begin{equation} \label{eq:confscalar}
\begin{aligned}
\phi(r) &   =  \frac{M \phi_{\infty}}{r-M} \\
&=\phi_{\infty} \sum_{j=1}^{\infty} \frac{M^j}{r^j}.
\end{aligned}
\end{equation}
It is easy to recognise that the metric is the same as the extreme RN one [i.e., expression \eqref{eq:rnmet} with $Q \to M$]. There exists an event horizon at $r=M$ where the scalar field diverges, similar to the JNW spacetime from Sec.~\ref{sec:jnw} (although the BBMB \emph{metric} is regular at the horizon).

The multipole moments for the BBMB spacetime have not been calculated before in the literature to the best of our knowledge, so we provide a derivation here using the procedure (i)--(iv) detailed in Sec.~\ref{sec:multmoms}. We begin by making the general observation that $\lambda = A(r)$ to find that the conformal Laplace equation \eqref{eq:multmomeqn} reads \cite{sm16} 
\begin{equation} \label{eq:bbmbm1}
0 = \tM''(\lambda) + \lambda^{-1} \tM'(\lambda) - \tfrac{1}{16} \lambda^{-2} \tM(\lambda).
\end{equation}
Equation \eqref{eq:bbmbm1} has the exact solution $\tM = c_{1} \lambda^{-1/4} + c_{2} \lambda^{1/4}$. Applying the boundary conditions \eqref{eq:massbc}, we find that the two constants are fixed as $c_{1} = -c_{2} = 1$. For the next step, we rescale the radial coordinate through
\begin{equation}
\bar{R} = \frac{1}{r - M},
\end{equation}
from which it is clear that the point at infinity is effectively mapped to $ \Lambda \to \lim_{r \to \infty} \bar{R}(r) = 0$. This makes it easy to pick a suitable conformal factor, viz.
\begin{equation}
\Omega = \bar{R}^2,
\end{equation}
for which it can be readily checked that the desired property, $\tilde{h}_{ij} = \Omega^2 h_{ij} = \delta_{ij}$, holds. Finally, we find the only non-zero mass moment is, as expected,
\begin{equation}
\begin{aligned}
M_{0} &= \lim_{\bar{R} \to \Lambda} \Omega^{-1/2} \tilde{\tau}_{M} \\
&= \lim_{\bar{R} \to 0} \bar{R}^{-1} \left(\frac{M \bar{R}}{\sqrt{1 + M \bar{R}}} \right) \\
&= M.
\end{aligned}
\end{equation}
For the scalar moments, we invert \eqref{eq:confscalar} to express $r$ as a function of $\phi$ from which we obtain, via equation \eqref{eq:multmomeqn}, 
\begin{equation} \label{eq:bbmbphieq}
0 = \tP''(\phi) + \frac{4 \left( \phi_{\infty} + \phi \right) \tP'(\phi) - \tP(\phi)}{4 \left( \phi_{\infty} + \phi\right)^{2}},
\end{equation}
which is similar to \eqref{eq:bbmbm1}. Solving equation \eqref{eq:bbmbphieq} subject to the boundary conditions \eqref{eq:scalarbc} and carrying out the same procedure as above, we find the non-zero moment
\begin{equation} \label{eq:phimombbmb}
\begin{aligned}
\Phi_{0} &= \lim_{\bar{R} \to \Lambda} \Omega^{-1/2} \tilde{\tau}_{\Phi} \\
&= \lim_{\bar{R} \to 0} \bar{R}^{-1} \left( \frac{\bar{R} M \phi_{\infty}}{\sqrt{1+ M \bar{R}}} \right) \\
&= M \phi_{\infty},
\end{aligned}
\end{equation}
as expected from the expansion \eqref{eq:confscalar}. At this point we note that there exists a charged generalisation of the BBMB solution \cite{bek74} that remarkably has \emph{the same metric} but with $M \to \sqrt{M^2 - Q^2}$ applied to the prefactor in the scalar potential \eqref{eq:confscalar} (see equation 1 in Ref.~\cite{bron78}). This again just renormalises $\Phi_0$ while introducing an electric monopole, $P_{0} = Q$, since $A_{t} \underset{r\to\infty}{\sim} Q/r$. In summary, the moments for the charged BBMB solution therefore read
\begin{equation} \label{eq:bbmbmoments}
M_{0} = M, \Phi_0 = \phi_{\infty}\sqrt{M^2 - Q^2} , P_{0} = Q.
\end{equation}

\section{Moments do not uniquely encapsulate spacetime structure} \label{sec:degen}

The properties of the three spacetimes considered in Sec.~\ref{sec:examples} and their moments are summarised in Table~\ref{tab:keyprops}. In particular, we see that it is, in principle, possible for two spacetimes to possess the same set of multipole moments and yet differ at the level of the line element once the theory of gravity departs from GR. The clearest example involves taking $Q \to 0$ in the charged JNW spacetime and adopting the areal radius, $\bar{r} = \sqrt{D(r)}$, to compare directly with the BBMB solution. Using expression \eqref{eq:dfunc} and the expansion in \eqref{eq:stmet}, it is easy to show that
\begin{equation}
    A_{\rm JNW}(\bar{r}) = 1 - \frac{2 M}{\bar{r}} + \mathcal{O}\left(\frac{1}{\bar{r}^3}\right).
\end{equation}
After an arbitrary renaming of the coordinate $\bar{r}$ we see that the above clearly differs from expression \eqref{eq:bbmb} as the latter contains a term proportional to $1/r^2$. At the same time, all moments coincide for these spacetimes provided we identify the constants through $n^2 = \kappa/(6+\kappa)$ (Tab.~\ref{tab:keyprops}).

By contrast, taking the charge to the extremal limit in the RN solution brings about the BBMB metric potential \eqref{eq:bbmb} but clearly the moments cannot match because the former spacetime is electrically charged while the latter need not be. This implies that not only do the moments not uniquely ``select'' a spacetime, the rank of the sourcing quantity (i.e., gravitational, electromagnetic, or scalar) also may not be discernable. This result is physically intuitive: all forms of energy gravitate in geometric theories but the \emph{origin} of the energy is, in some sense, invisible at the level of the metric (see Ref.~\cite{suv21} for a discussion on such ``gravitational doppelg{\"a}ngers'').

Given that all three of the geometries have a Schwarzschild limit, the result we have reached may be rephrased in terms of Birkhoff's theorem (or a lackthereof in modified gravity). Once additional hairs are permitted to exist, the way in which the exterior spacetime responds to their presence depends on the field equations. In cases with non-minimal couplings [such as the action \eqref{eq:BBMBaction}], the spacetime-backreaction of non-tensorial fields may be amplified in the vicinity of a compact object even if the fields extracted at conformal infinity are identical. Since the moments are only defined by such an extraction, information regarding any intermediate behaviour is lost. This is why it is possible for two distinct spacetimes to possess the same moments, and why it is not possible in Newtonian theory or GR due to uniqueness theorems applying to the relevant differential equations (cf. Refs.~\cite{amm16,good18}). As this applies even in the static limit, it must also hold for more general stationary cases where rotation is considered (see Ref.~\cite{yaz25} for a discussion on uniqueness theorems in scalar-tensor theories).

\subsection{Application to black hole shadows} \label{sec:deflection}

As a simple application, we demonstrate explicitly that some astrophysical quantities may not be expressed uniquely in terms of moments. For this purpose, we consider the deflection of photons in the spacetime described by expression \eqref{eq:genline} in the context of geometric optics. Calculating photon trajectories is relevant, for instance, in the determination of black hole shadows \cite{sch92} to theoretically compare models with astrophysical observations made by the Event Horizon Telescope (EHT; \cite{eht1}).

From the particle Euler-Lagrange equations,
\begin{equation} \label{eq:eulerlagrange}
0 = \frac{d}{d \sigma} \left( \frac{\partial \mathcal{L}}{\partial \dot{x}^{\mu}} \right) -  \frac{\partial \mathcal{L}}{\partial x^{\mu}},
\end{equation}
where $\mathcal{L} = \tfrac{1}{2}g_{\mu \nu} \dot{x}^{\mu} \dot{x}^{\nu}$ for 4-momenta $\dot{\boldsymbol{x}}$ and affine parameter $\sigma$, it is straightforward to show that the critical impact parameter is given by \cite{per22}
\begin{equation} \label{eq:genbcrit}
b_{\rm cr}^2 = \left[ \frac{D(r)}{A(r)} \right]_{{r = r_{\rm ph}}},
\end{equation}
where $r_{\rm ph}$ denotes the radius of the photon sphere and the metric $g_{\mu \nu}$ is that defined by expression \eqref{eq:genline}. Physically, $b_{\rm cr}$ denotes the impact parameter separating captured and flyby orbits of incident light rays. For a static, spherically-symmetric black hole, it represents the radius of the ``shadow'' seen by a distant observer.

For the RN spacetime, we have the classical result \cite{eir02}
\begin{equation} \label{eq:rnbcrit}
b^{\rm RN}_{\rm cr} = \sqrt{\frac{M_0 \sqrt{(9 M_0^2 - 8 P_0^2)^3} + 8 P_0^4 - 36 P_0^2 M_0^2 + 27 M_0^4}{2 (M_0^2 - P_0^2)}},
\end{equation}
where, importantly, we have re-expressed constants in terms of the relevant monopole moments via expressions \eqref{eq:rnmoments}. In GR, this result is unique in the sense that if one could measure $M_{0}$ and $P_{0}$ together with $b_{\rm cr}$ and found a mismatch with expression \eqref{eq:rnbcrit}, an exotic object or astrophysical distortions to the spacetime would be required in the static limit (see, e.g., Refs.~\cite{gp21,gp23}).

\begin{figure}
\begin{center}
\includegraphics[width=0.487\textwidth]{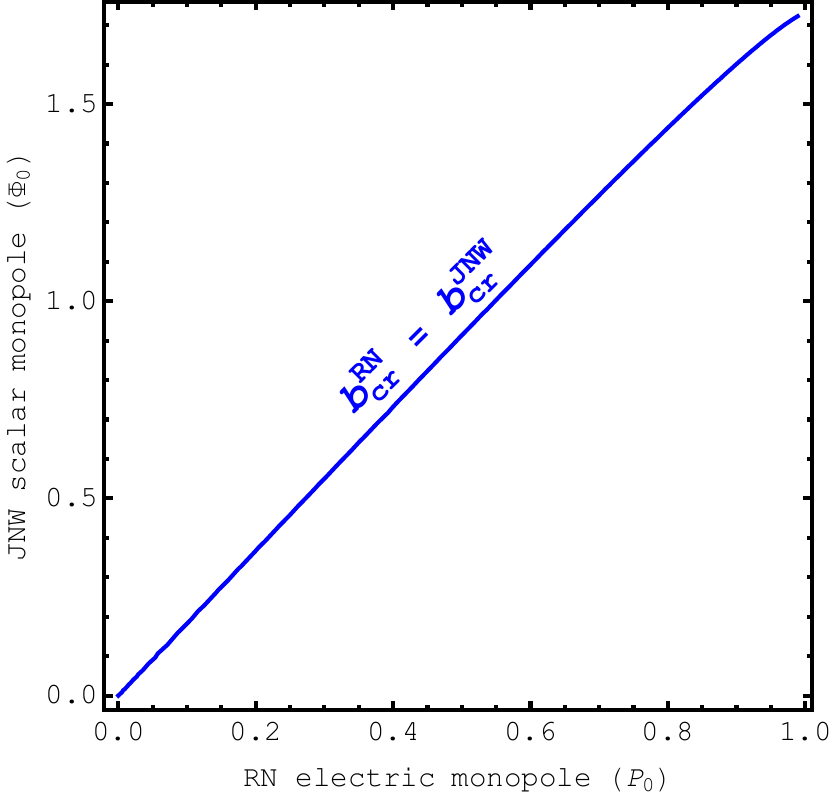}
\caption{Depiction of equality between the critical impact parameters for the RN and JNW spacetimes, $b^{\rm RN}_{\rm cr} = b^{\rm JNW}_{\rm cr}$, as a function of RN electric monopole ($P_{0}$) and JNW scalar monopole ($\Phi_0$). Units such that $M_0=1$ are used.
\label{fig:crits}}
\end{center}
\end{figure}

By contrast, for $Q=0$, we have for the (exotic object) JNW spacetime that $r_{\rm ph} = M(2+1/n)$ and thus \cite{tur24}
\begin{equation} \label{eq:jnwbcrit}
b^{\rm JNW}_{\rm cr} = M_0 \sqrt{3 - \frac{\Phi_0^2}{M_0^2}} \left( \frac{2M_0 + \sqrt{M_0^2 + \Phi_0^2}} {2 M_0 - \sqrt{M_0^2 + \Phi_0^2}}\right)^{\frac{M_0}{\sqrt{M_0^2 + \Phi_0^2}}}.
\end{equation}
In the limit of vanishing scalar hair, $\Phi_0 \to 0$, we recover the Schwarzschild expression, $b^{\rm Sch}_{\rm cr} = 3 \sqrt{3} M_0$. 
Even in the absence of charge for the JNW spacetime, $b^{\rm RN}_{\rm cr}$ and $b^{\rm JNW}_{\rm cr}$ can be made to match numerically for any given $P_{0} \leq M$. 
This is demonstrated in Figure~\ref{fig:crits}, showing that there exists a $\Phi_0$, for any given $P_0$, such that $b^{\rm RN}_{\rm cr} = b^{\rm JNW}_{\rm cr}$. 
For instance, taking $P_0 = 0.5 M_0$ yields the solution $\Phi_0 \approx 0.914 M_{0}$ where both impact parameters take value $b_{\rm cr} \approx 4.97 M_0$. 
There is thus a degeneracy between electromagnetic and scalar moments in the context of shadows. 
This is further exemplified by considering the BBMB metric with its soft hair, for which we find
\begin{equation}
b^{\rm BBMB}_{\rm cr} = 4 M_0,
\end{equation}
which is a factor $\sim 1.3$ smaller than the Schwarzschild value for any given mass and electric monopole moments. 

The implication of this demonstration is that due to the non-uniqueness of moments across theories, they alone are insufficient to express physical quantities (e.g. shadow radii) \emph{when the theory of gravity is not held fixed}.

\section{Conclusions} \label{sec:discussion}

In this paper, we explored the construction of multipole moments in various classical theories of gravity. In any given theory, these can be computed by appealing to analogies with the Poisson equation of the Newtonian framework, as in the seminal works of Geroch and Hansen \cite{ger70a,ger70,han74}; the results for three simple examples are given in Tab.~\ref{tab:keyprops}. The key point is that a triplet of monopole moments (in the scalar, electromagnetic, and tensorial sectors) can match between two spacetimes even if their metric structures are different. In a static spacetime, these are the only moments that can be non-vanishing. The implication is that, unless the theory of gravity is fixed \emph{a priori}, knowledge of the moments alone is insufficient to uniquely categorise physical observables.

While we focussed on the simplified case of shadows cast by static, spherically-symmetric objects in Sec.~\ref{sec:deflection} to demonstrate the above point, the result has implications for other astrophysical phenomena. For example, tight correlations between the moments of inertia, Love numbers, and quadrupole moments found for a number of compact object spacetimes (e.g., Refs.~\cite{yagi13,Pappas:2013naa,Yagi:2014bxa,yagi14,mart14,pap19}) provide a promising tool to extract the maximum amount of information about their structure given a set of multimessenger observations. If different spacetimes can possess the same moments however---as we have shown---it is clear that such relations are implicitly sensitive to the theory under consideration. The notion of ``universal relations'' thus requires an implicit ``in-GR'' prefix and may not apply to multiple theories simultaneously. How severe this restriction is, together with moment-mappings for gravitational waves and geodesic motions in other theories \cite{ryan95,ryan97}, will be investigated elsewhere.

\begin{acknowledgments}
We thank Kostas Glampedakis for discussions. This paper is part of a project that has received funding from the European Union's Horizon Europe Research and innovation programme under Grant Agreement No 101131928. AGS acknowledges funding from the European Union's Horizon MSCA-2022 research and innovation programme ``EinsteinWaves'' under grant agreement No. 101131233.
\end{acknowledgments}


%


\appendix

\section{Non-uniqueness within a fixed theory} \label{sec:Appendix}

{In addition to the results discussed in the main text, it is worth pointing out that even \emph{within} a given theory a set of moments may fail to uniquely capture vacuum structure. 
This can be demonstrated -- again by counterexample -- by simply noting that any free parameters appearing within the action of a theory will typically manifest within solutions. 
As such, since only monopole moments can be non-zero in static and spherically-symmetric cases, if the (uncharged) theory hosts at least two coupling parameters then it is clear that the two available moments cannot contain enough information to catalogue solutions (i.e., the mass and two innate hairs).}

{Such a restriction is not necessary however, as even some theories with only a single, short-range parameter may be degenerate in the sense of moments. 
In fact, even some theories with \emph{zero} free action parameters may be degenerate. 
One obvious example is the pure $R^2$ theory. 
Any metric with vanishing Ricci scalar is a solution in these theories \cite{suv21} and thus, since there are (at least) two free metric functions $A$ and $B$ in expression \eqref{eq:genline}, an arbitrary number of free parameters can appear in a static solution.
A small set of monopole moments clearly cannot encode such near-field details in general.}

{As a less contrived example, consider the hybrid Gauss-Bonnet-Horndeski theory introduced by \citet{lupang20}, whose action reads
\begin{equation} \label{eq:hornaction}
\begin{aligned}
S=&\int d^4x\sqrt{-g}\Big\{R
+\alpha\big[\phi \mathcal{G} +4G^{\mu\nu}\nabla_\mu\phi\nabla_\nu\phi \\
&-2 \chi Re^{-2\phi}-4\nabla_{\mu}\phi\nabla^{\mu}\phi (\nabla_{\nu} \nabla^{\nu} \phi)\\
&+2(\nabla_{\mu}\phi\nabla^{\mu} \phi)^2-12\chi e^{-2\phi}\nabla_{\mu}\phi\nabla^{\mu} \phi \\
&-6 \chi^2 e^{-4\phi}\big]\Big\}.
\end{aligned}
\end{equation}
Here, $G^{\mu \nu}$ is the Einstein tensor, $\mathcal{G} = R_{\mu\nu\rho\sigma}R^{\mu\nu\rho\sigma}-
4R_{\mu\nu}R^{\mu\nu}+R^2$ is the Gauss-Bonnet invariant, and GR is recovered by setting the coupling parameter, $\alpha$, to zero. 
The other parameter, $\chi$, is inessential and can effectively be absorbed by field redefinitions \cite{bab23}.
There exists an exact solution in this theory with metric taking the form of expression \eqref{eq:genline} with $B = 1/A$ and $D = r^2$, such that
\begin{equation}  \label{eq:hornafn}
\begin{aligned}
    A(r) &= 1 + \frac{r^2}{2\alpha} \left( 1 - \sqrt{1+\frac{8 \alpha M}{r^3}} \right) \\
    &= 1 - \frac{2 M}{r} + \frac{4 \alpha M^2}{r^4} + \mathcal{O}\left(\frac{1}{r^{7}}\right).
    \end{aligned}
\end{equation}
The associated scalar field can be written in full in terms of integrals of $A$, though has the convenient expansion
\begin{equation} \label{eq:hornscalar}
    \phi(r) = \frac{M}{r} + \frac{3 M^2}{4 r^2} + \frac{5 M^3}{6 r^3} + \mathcal{O}\left(\frac{1}{r^{4}}\right).
\end{equation}
What is interesting is that the parameter $\alpha$ appears \emph{only} at order $1/r^4$ in the potential \eqref{eq:hornafn} and not at all within the scalar field \eqref{eq:hornscalar}. 
Though not shown rigorously here, it is clear that the mass- and scalar-monopole moments for this solution thus both take value $M$ (up to sign),
\begin{equation} \label{eq:hornmoments}
    M_{0} = M, \qquad \Phi_{0} = M.
\end{equation}
Since the parameter $\alpha$ does not appear in expressions \eqref{eq:hornmoments}, this implies that there are infinitely many solutions within the theory described by \eqref{eq:hornaction} corresponding to a given set of moments. 
Care must be exercised therefore in describing ``universal relations'' involving multipole moments within this or similar theories.}

\end{document}